\documentclass[pdftex,twocolumn,epjc3]{svjour3}          

\RequirePackage[T1]{fontenc}

\smartqed  

\RequirePackage{graphicx}
\RequirePackage{mathptmx}      
\RequirePackage{flushend}
\RequirePackage[numbers,sort&compress]{natbib}
\RequirePackage[colorlinks,citecolor=blue,urlcolor=blue,linkcolor=blue]{hyperref}
\usepackage{amsmath}
\journalname{Eur. Phys. J. C}

\newcommand {\sNN}[1]{$\sqrt{s_{\rm NN}} = #1$}

\newcommand {\pp}{$p$+$p$}
\newcommand {\ee}{$e^{+}e^{-}$}
\newcommand {\pt}{p_{\rm T}}

\newcommand {\gevc}{GeV/$c$}

\begin{document}

\title{Contribution of coherent electron production to measurements of heavy-flavor decayed electrons in heavy-ion collisions}


\author{Shenghui Zhang \thanksref{addr, t1}
\and Rongrong Ma \thanksref{addr2}
\and Yuanjing Ji \thanksref{addr3}
\and Zebo Tang \thanksref{addr1}
\and Qian Yang \thanksref{addr4}
\and Yifei Zhang \thanksref{addr1}
\and Wangwei Zha\thanksref{addr1, t2}
}

\thankstext[a]{t1}{e-mail: zhangshh08@cqu.edu.cn}
\thankstext[b]{t2}{e-mail: first@ustc.edu.cn (corresponding author)}

\institute{Department of Physics, Chongqing Key Laboratory for Strongly Coupled Physics, Chongqing University, Chongqing 401331, P.R. China\label{addr}
\and Department of Modern Physics, University of Science and Technology of China, Hefei 230026, P.R. China\label{addr1}
\and Physics Department, Brookhaven National Laboratory, Upton, New York 11973, USA\label{addr2}
\and Nuclear Science Division, Lawrence Berkeley National Laboratory, Berkeley, California 94720, USA\label{addr3}
\and Institute of Frontier and Interdisciplinary Science, Shandong University, Qingdao 266237, P.R. China\label{addr4}
}

\date{\today}


\maketitle

\begin{abstract}
Heavy quarks, produced at early stages of heavy-ion collisions, are an excellent probe of the Quark-Gluon Plasma (QGP) also created in these collisions. Electrons from open heavy-flavor hadron decays (HFE) are good proxies for heavy quarks, and have been measured extensively in the last two decades to study QGP properties. These measurements are traditionally carried out by subtracting all known background sources from the inclusive electron sample. More recently, a significant enhancement of $e^+e^-$ pair production at very low transverse momenta was observed in peripheral heavy-ion collisions. The production characteristics is consistent with coherent photon–photon interactions, which should also constitute a background source to the HFE measurements. In this article, we provide theoretical predictions for the contribution of coherent electron production to HFE as a function of transverse momentum, centrality and collision energy in Au+Au and Pb+Pb collisions.

\end{abstract}

\section{Introduction}
Ultra-relativistic heavy-ion collisions at the Relativistic Heavy-Ion Collider (RHIC) and the Large Hadron Collider (LHC) are a unique laboratory for studying the properties of the strong interaction described by Quantum Chromodynamics (QCD). Numerous measurements from RHIC and LHC experiments have provided strong evidence that a novel state of strongly-interacting matter, called the Quark-Gluon Plasma (QGP) and composed of deconfined quarks and gluons ~\cite{ref:QGP1, ref:QGP2}, is created in these collisions ~\cite{BRAHMS:white:paper,STAR:white:paper,PHENIX:white:paper,PHOBOS:white:paper,LHC,RL}.


Due to their large masses, heavy quarks, {\it i.e.} charm and bottom quarks, are primarily produced in hard partonic scatterings at the early stage of heavy-ion collisions before the formation of the QGP~\cite{hfreview}. Subsequently, they traverse the QGP throughout its evolution and encode fundamental QGP properties, such as the heavy-quark transport coefficients \cite{PhysRevC.99.054907}. The strong interaction between heavy quarks and QGP constituents is expected to result in substantial energy loss for heavy quarks, which can manifest experimentally as a reduction of the production rate for heavy-flavor hadrons at a given momentum \cite{Andronic:2015wma}. For example, significant suppression of the charm meson yields at large transverse momenta ($\pt$) has been observed in heavy-ion collisions, compared to those in \pp\ collisions, at both RHIC and the LHC~\cite{hfreview1, hfreview2, D0_STAR, D0_ALICE, D0_ALICE1, D0_CMS}.

Electrons \footnote{Unless specified otherwise, electrons referred to here include both electrons and positrons and results are presented as $\frac{e^++e^-}{2}$.} from semileptonic decays of heavy-flavor hadrons (HFE) can serve as good proxies for measuring heavy quarks. Although one does not have direct access to the kinematics of heavy-flavor hadrons through HFE, their higher branching ratios compared to the hadronic decay channels and the experimental capability to trigger on high-$\pt$ electrons for enhancing statistics make HFE a very good tool to study heavy quark production in heavy-ion collisions~\cite{STAR:NPE, STAR:2023qfk, 2023138071, PHENIX:NPE, ALICE:NPE1, ALICE:NPE2}. Experimentally, the HFE yield is usually obtained by subtracting all known background sources, such as photon conversions, Dalitz decays of $\pi^{0}$ and $\eta$ mesons, Drell-Yan process, etc from the inclusive electron sample \cite{Zhang:2021iqp, STAR:2023qfk, STAR:ppNPE, STAR:NPE2}. It is worth noting that this method relies on the precise knowledge of background sources, which could evolve over time. 

In the last few years, significant enhancements of dilepton pair production at low $\pt$ above known hadronic sources are observed in Au+Au and Pb+Pb collisions with large impact parameters at the center-of-mass energy per nucleon-nucleon pair of \sNN{200} GeV and 5.02 TeV by the STAR collaboration at RHIC, the ALICE and ATLAS collaborations at LHC, respectively~\cite{STARdilepton, dielectron_ALICE, PhysRevLett.121.212301}. These enhancements are incompatible with QGP thermal radiation or in-medium $\rho$ broadening~\cite{STARdilepton, dielectron_ALICE}. The concentration of the excess yields at low $\pt$ inspired the explanation of this contribution as resulting from coherent photon-photon interactions that occur in addition to the hadronic nucleus-nucleus collision. These photons are generated by the large electrical charge of the two incoming nuclei ($Z \sim 80$). Theoretical calculations of the coherent process~\cite{photondilepton, photondilepton1, photondilepton2, ZHA2020135089, 7-20230074, PhysRevD.104.056011, PhysRevD.106.034025, PhysRevD.102.094013, KLUSEKGAWENDA2021136114} indeed provided quantitative description of data. A natural question prompted by the new knowledge gained from these developments is how much the coherently produced electrons will contribute to the HFE measurements. In this article, we present calculations of the contributions of electrons from coherent photon-photon interactions to HFE measurements in Au+Au and Pb+Pb collisions at $\sqrt{s_{\rm NN}} $ = 7.7 GeV - 5.02 TeV.


\section{Analysis setup}
\subsection{Coherent photon-photon interaction}

In heavy-ion collisions, the large fluxes of photons, induced by highly-charged colliding nuclei traveling close to the speed of light, can interact with each other coherently, and produce \ee\ pairs through the Breit-Wheeler process~\cite{PhysRevD.4.1532, BW, PhysRevC.107.044906, EPJAZ}. With the formalism of the equivalent photon approximation~\cite{CREE}, the coherent photon–photon interaction can be factorized into a semiclassical and a quantum-mechanical part. The semiclassical part calculates the distribution of quasi-real photons induced by the incoming ions, while the quantum-mechanical part describes the interaction cross section. The production of \ee\ pairs can therefore be expressed as~\cite{CREE, PhysRevC.47.2308}:
\begin{eqnarray}
\begin{aligned}
&\sigma (A + A \rightarrow A + A + e^{+}e^{-})\\
&=\int d\omega_{1}d\omega_{2}n(\omega_{1})n(\omega_{2})\sigma_{\gamma \gamma}(W),
\label{eq1}
\end{aligned}
\end{eqnarray}
where $A$ stands for the incoming nucleus, $\omega_{1}$ and $\omega_{2}$ are the energies of photons emitted by the two colliding beams,  $n(\omega)$ is the photon flux at energy $\omega$, $W$ is the \ee\ pair invariant mass, and $\sigma_{\gamma \gamma}(W) = \sigma[\gamma \gamma \rightarrow e^{+}e^{-}(W)]$ is the cross section for pair production. The pair mass ($W$) and pair rapidity ($y$) can be determined using photon energies:
\begin{eqnarray}
\label{eq2}
W=\sqrt{4\omega_1\omega_2}
\end{eqnarray}
and
\begin{eqnarray}
\label{eq3}
y=\frac{1}{2}{\rm ln}\frac{\omega_1}{\omega_2}.
\end{eqnarray}

\begin{figure}[!htb]
\centering
\includegraphics[width=\linewidth]{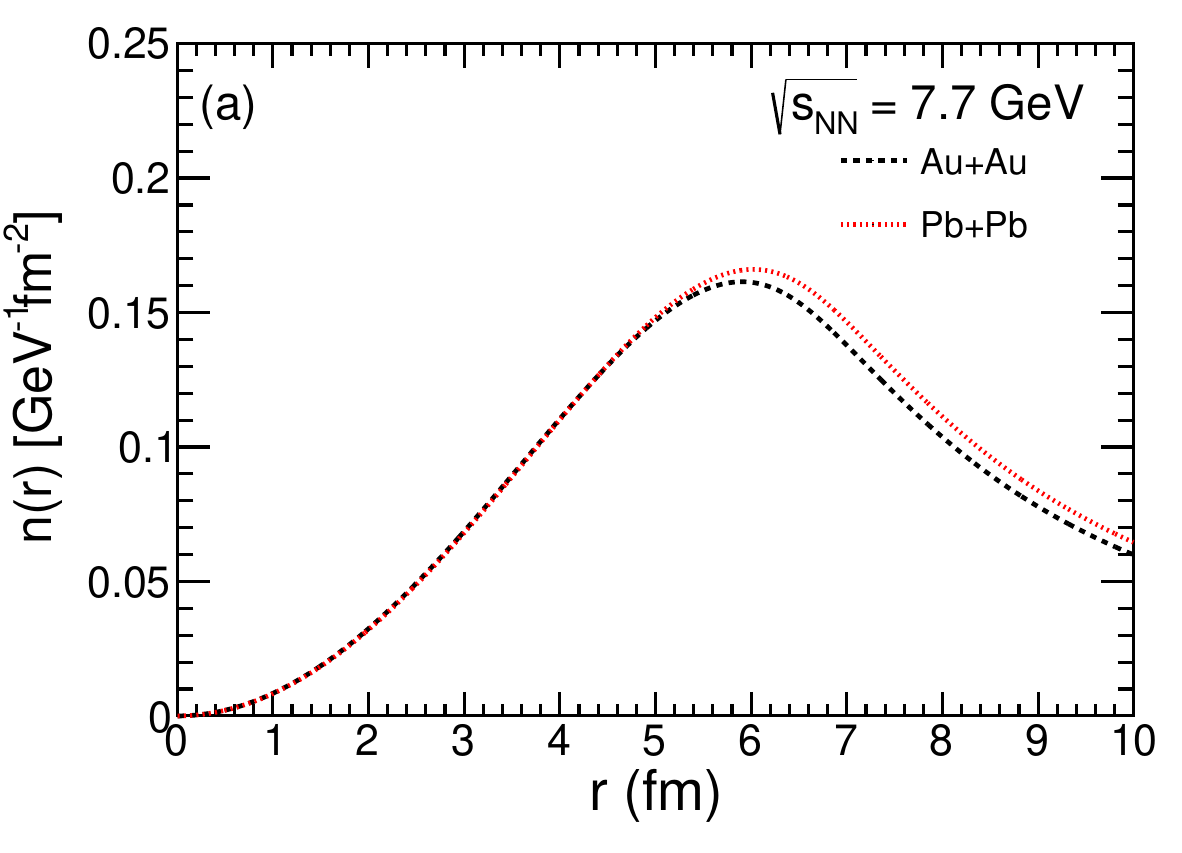}
\includegraphics[width=\linewidth]{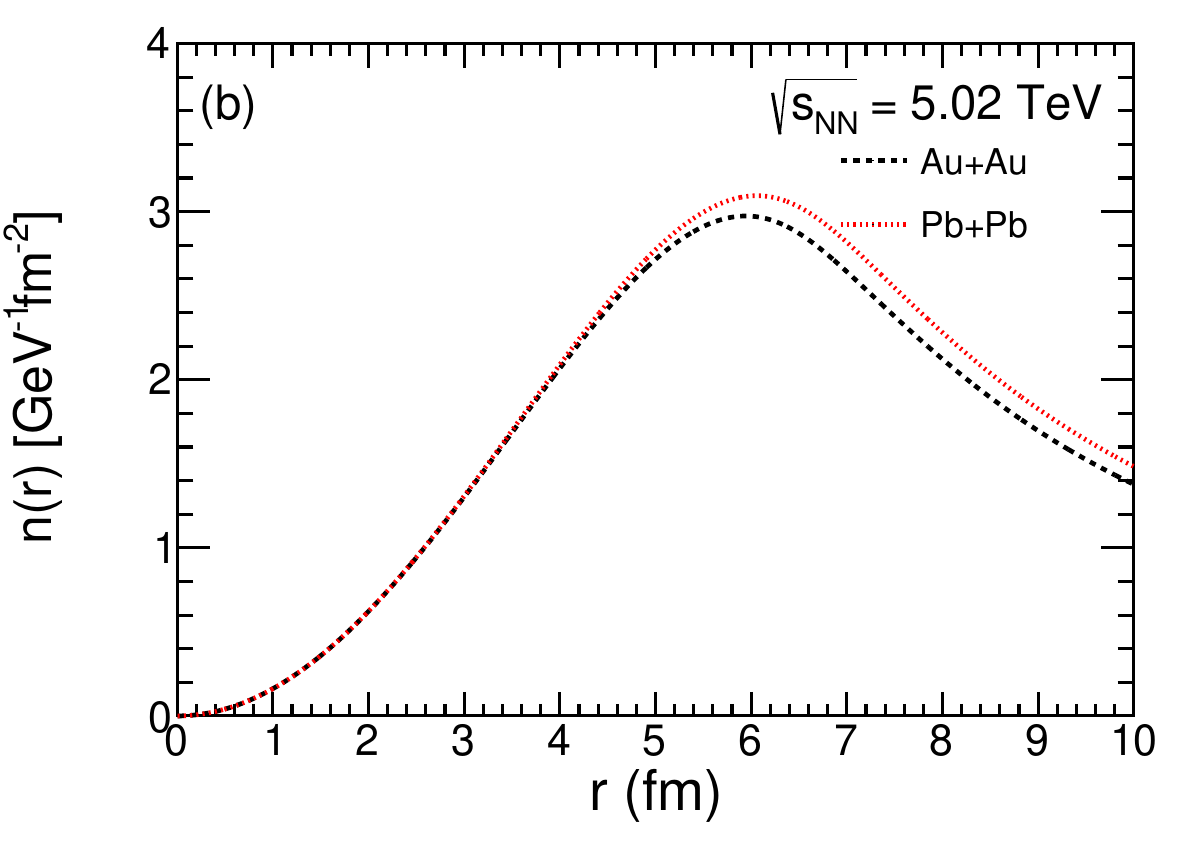}
\vspace{-2ex}
\caption{Photon flux as a function of distance $r$ from the center of emitting Au and Pb nucleus in collisions of \sNN{7.7} GeV (a) and 5.02 TeV (b).}
\label{fig1new}
\end{figure}

The photon flux with energy $\omega$ and at distant $\vec{r}$ from the center of the emitting nucleus can be modelled using the Weizs\"acker-Williams method~\cite{CREE}:
\begin{eqnarray}
\label{eq4}
\begin{aligned}
  & n(\omega,\vec{r}) = \frac{4Z^{2}\alpha}{\omega} \bigg | \int \frac{d^{2}\vec{q}_{\bot}}{(2\pi)^{2}}\vec{q}_{\bot} \frac{F(\vec{q})}{\vec{q}^{2}} e^{i\vec{q}_{\bot} \cdot \vec{r}} \bigg |^{2}\\
  & \vec{q} = (\vec{q}_{\bot},\frac{\omega}{\gamma})
\end{aligned}
\end{eqnarray}
where $\alpha=1/137$ is the electromagnetic coupling constant, $Z$ is the nuclear charge number, $\vec{q}_{\bot}$ is the transverse momentum of the photon, $\gamma$ is Lorentz factor. The form factor $F(\vec{q})$, which carries the information about the charge distribution inside the nucleus, can be obtained by performing a Fourier transformation to the nuclear charge density $\rho(\vec{r})$:
\begin{eqnarray}
\label{eq5}
 F(\vec{q})=\int {\rm d}^3\vec{r}\rho(\vec{r})e^{i\vec{q}\cdot \vec{r}}.
\end{eqnarray}
For a spherical nucleus, such as Au and Pb, the form factor can be expressed as follows:
\begin{equation}
  F(q)=\frac{4\pi}{q}\int {\rm d}r r\rho(r){\rm sin}(qr).
\label{eq6}
\end{equation}
In this work, we employ the Woods-Saxon form to model the charge density for a spherical nucleus:
   \begin{equation}
   \rho(r)=\frac{\rho^{0}}{1+\exp[(r-R_{\rm{WS}})/d]},
  \label{eq7}
  \end{equation}
  where $\rho^0$ is the normalization factor and also denotes the charge density at the center of the nucleus (Au: $0.16$ $fm^{-3}$, Pb: $0.17$ $fm^{-3}$), the radius $R_{\rm{WS}}$ (Au: 6.38 $fm$, Pb: 6.62 $fm$) and skin depth $d$ (Au: 0.535 $fm$, Pb: 0.546 $fm$) are based on fits to electron scattering data~\cite{10.1063/1.2995107,DEJAGER1974479}. Previous studies have shown that the difference in the photon flux between using Woods-Saxon and point-like forms is negligible for $r>>R_{\rm{WS}}$~\cite{photondilepton}. Therefore, we use the photon flux calculated with Woods-Saxon form factor for $r < 10$ $fm$; while at  $r > 10$ $fm$, the point-like charge distribution is employed. For the point-like charge distribution, the photon flux is given by a simple formula:
    \begin{equation}
    n(\omega,r) = \frac{Z^{2}\alpha}{\pi^{2}\omega r^{2}}x^{2}K_{1}^{2}(x),
         \label{eq8}
    \end{equation}
where $x=\omega r/\gamma$, and $K_{1}$ is the modified Bessel function. Figure~\ref{fig1new} shows the photon flux with Woods-Saxon form as a function of distance $r$ from the center of emitting nucleus in Au+Au and Pb+Pb collisions at \sNN{7.7} GeV and 5.02 TeV. A distinct peak structure is seen around $r\sim 6$ fm. Furthermore, the photon flux is larger for the Pb nucleus than for the Au nucleus due to the larger electric charge carried by the Pb nucleus, and larger at higher collision energy.

\begin{figure*}[!htb]
\centering
\includegraphics[width=0.75\linewidth]{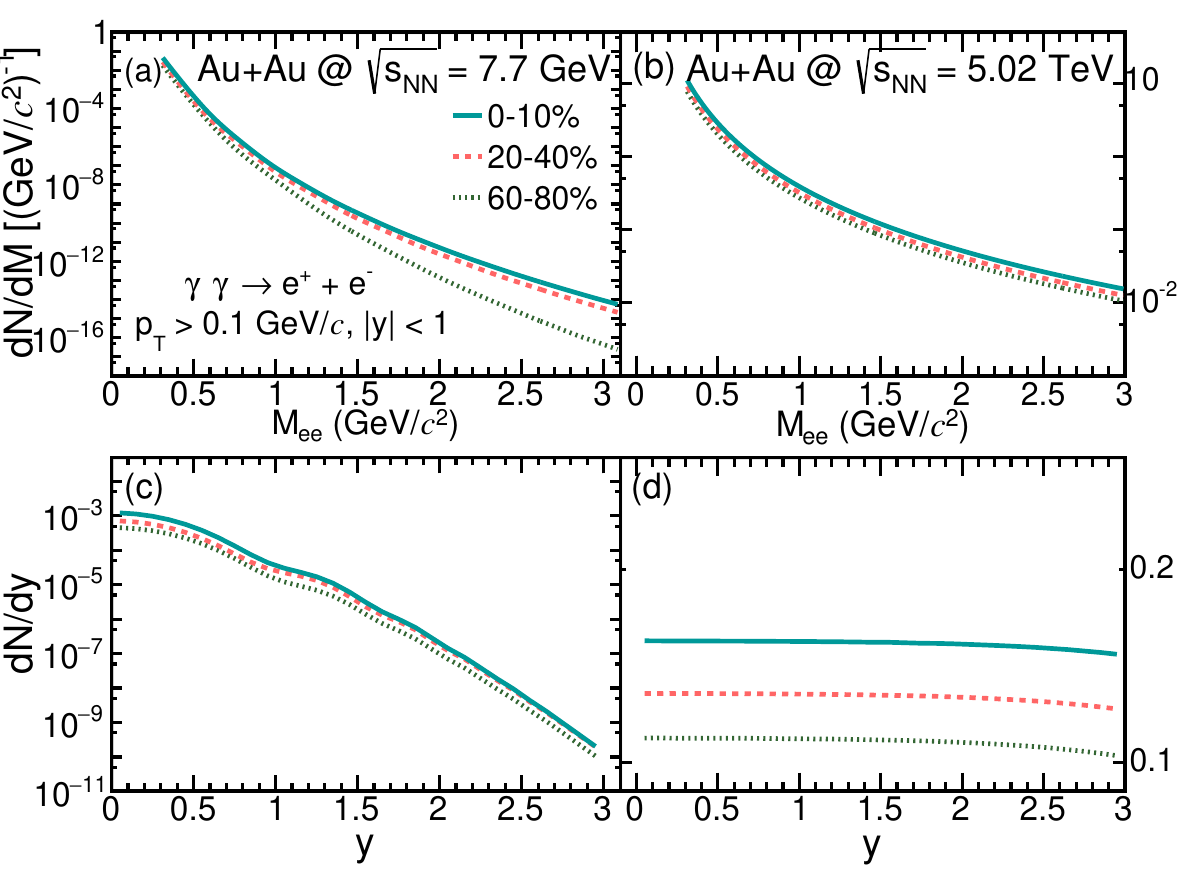}
\vspace{-2ex}
\caption{(a) Invariant mass spectra of the coherently produced \ee\ pairs for different centrality classes of Au+Au collisions at \sNN{7.7} GeV. (b) Same as (a) except that it is for 5.02 TeV. (c) Rapidity distributions of the coherently produced \ee\ pairs for different centrality classes of Au+Au collisions at \sNN{7.7} GeV. (d) Same as (c) except that it is for 5.02 TeV.}
\label{fig1}
\end{figure*}

The cross section for producing a \ee\ pair with mass $W$ can be determined by the Breit-Wheeler formula~\cite{PhysRevD.4.1532}
\begin{eqnarray}
\begin{aligned}
&\sigma_{\gamma \gamma}(W)\\
&=\frac{4\pi \alpha^{2}}{W^{2}} [(2+\frac{8m_{e}^{2}}{W^{2}} - \frac{16m_{e}^{4}}{W^{4}})\times \text{ln}(\frac{W+\sqrt{W^{2}-4m_{e}^{2}}}{2m_{e}})\\
&\quad -\sqrt{1-\frac{4m_{e}^{2}}{W^{2}}}(1+\frac{4m_{e}^{2}}{W^{2}})],
\label{eq9}
\end{aligned}
\end{eqnarray}
where $m_{e}$ is the electron rest mass. The angular distribution of these pairs is given by
 \begin{equation}
  G(\theta) = 2 + 4(1-\frac{4m_{e}^{2}}{W^{2}})\frac{(1-\frac{4m_{e}^{2}}{W^{2}})\text{sin}^{2}(\theta)\text{cos}^{2}(\theta)+\frac{4m_{e}^{2}}{W^{2}}}{(1-(1-\frac{4m_{e}^{2}}{W^{2}})\text{cos}^{2}(\theta))^{2}},
  \label{eq10}
  \end{equation}
 where $\theta$ is the angle between the beam direction and the momentum of the electron in the electron-positron pair center-of-mass frame. Here, the effect of finite, but small photon $p_{\rm T}$ on the angular distribution is neglected. 

\begin{table}[htbp]
  \caption{\label{tab1}
Values of $b_{\rm min}$ and $b_{\rm max}$ corresponding to 60-80\% centrality class for Au+Au and Pb+Pb collisions at the different energies.
  }
{
\begin{tabular}{c|c|c}
Collision energy & Au+Au & Pb+Pb  \\
& ($b_{\rm min}$, $b_{\rm max}$) fm & ($b_{\rm min}$, $b_{\rm max}$) fm \\
\hline
7.7 GeV  & (11.26, 13.01) & (11.66, 13.47)\\   
19.6 GeV & (11.28, 13.04) & (11.68, 13.49)\\
27 GeV & (11.29, 13.05) & (11.69, 13.50) \\
39 GeV & (11.31, 13.06) & (11.70, 13.52) \\
54.4 GeV & (11.32, 13.08) & (11.72, 13.54) \\
62.4 GeV & (11.34, 13.10) & (11.73, 13.55)\\
130 GeV & (11.38, 13.15) & (11.78, 13.61) \\
200 GeV & (11.41, 13.18) & (11.80, 13.64) \\
2.76 TeV & (11.58, 13.38) & (11.98, 13.84) \\
5.02 TeV & (11.62, 13.43) & (12.03, 13.89) \\
\end{tabular}
}
\end{table}

With the convolution of photon flux and $\gamma \gamma \rightarrow e^{+}e^{-}$ cross section, the probability to produce a \ee\ pair through coherent photon-photon interaction in a heavy-ion collision with impact parameter $b$ can be given by:
  \begin{equation}
  P(W,y,b)=\frac{W}{2}\int d^{2}r_{1}n(\omega_{1},r_{1})n(\omega_{2},|\vec{b} - \vec{r}_{1}|)\sigma_{\gamma \gamma}(W).
  \label{eq11}
  \end{equation}
  Consequently, the invariant yield for coherently produced electrons in heavy-ion collisions of a selected centrality class can be written as:

\begin{equation}
  \frac{d^{2}N}{dWdy}=\frac{\int^{b_{\rm{max}}}_{b_{\rm{min}}} d^{2}b P(W,y,b) \times P_{H}(\vec{b})}{\int^{b_{\rm{max}}}_{b_{\rm{min}}} d^{2}b P_{H}(\vec{b})},
  \label{eq12}
\end{equation}
  with the hadronic interaction probability at $\vec{b}$
      \begin{equation}
  P_{H}(\vec{b}) = 1 - \text{exp}[-\sigma_{\text{NN}} \int d^{2}r T_{A}(\vec{r})T_{A}(\vec{r}-\vec{b})],
  \label{eq13}
  \end{equation}
  where $\sigma_{\text{NN}}$ is the inelastic nucleon-nucleon cross section,  which is dependent on collision energy, and $T_{A}(\vec{r})$ is the nuclear thickness function defined as
        \begin{equation}
T_{A}(\vec{r}) = \int dz \rho(\vec{r},z)dz.
  \label{eq14}
  \end{equation}
 $b_{\rm{min}}$ and $b_{\rm{max}}$ in Eq.~\ref{eq12} are the minimum and maximum impact parameters for the selected centrality class, and can be determined from the Monte Carlo Glauber simulation~\cite{LOIZIDES201513}.  Here, the term ``centrality" is used to represent the geometry of a heavy-ion collision. Peripheral (central) collisions correspond to those with large (small) impact parameters and thus small (large) nuclear overlap. Table~\ref{tab1} shows $b_{\rm{min}}$ and $b_{\rm{max}}$ values used for 60-80\% Au+Au and Pb+Pb collisions at different collision energies. Due to the larger size of the Pb nucleus, the impact parameter values are slightly larger for Pb+Pb than for Au+Au collisions. They also increase with increasing collision energy. As discussed in Ref.~\cite{photondilepton}, the impact of violent hadronic interactions, occurring in the overlap region, on photoproduction is negligible.

The resulting invariant mass spectra $dN/dM$ and rapidity distributions $dN/dy$ of the coherently produced \ee\ pairs are shown in Fig.~\ref{fig1} for different centrality classes of Au+Au collisions at \sNN{7.7} GeV and 5.02 TeV. The following kinematic cuts are applied: electron $p_{\rm T} >$ 0.1 GeV/$c$ and $|y| <$ 1. The production yield increases significantly with increasing collision energy, and also slightly from peripheral to central collisions. Furthermore, a much stronger rapidity dependence is seen at 7.7 GeV compared to that at 5.02 TeV. It is worth noting that there are no coherent \ee\ pairs produced below $M_{e^+e^-} <$ 0.3 GeV/$c^2$ due to the minimum electron $\pt$ cut of 0.1 GeV/$c$ used in this work. Therefore, these coherently produced electrons can't be rejected by the typical invariant mass cut of $M_{e^+e^-} <$ 0.1 GeV/$c^{2}$ employed in HFE measurements for removing photonic electron background~\cite{STAR:ppNPE, STAR:NPE2, STAR:2023qfk, ALICE:NPE, ALICE:NPE1, ALICE:NPE2}. 

It is noteworthy that calculations using the same framework as in this paper have been compared with experimental results, and they can describe the production of \ee\ and $\mu^{+}\mu^{-}$ pairs at very low $\pt$, commonly attributed to coherent process, quite well~\cite{STARdilepton, ZHA2020135089}. 

\subsection{Open heavy-flavor hadron decayed electrons}
For the HFE production within $|y| < 1$ in heavy-ion collisions at different energies, they are estimated by scaling the HFE yields in \pp\ collisions of corresponding energies with the average number of binary nucleon-nucleon collisions ($N_{\rm coll}$)~\cite{LOIZIDES201513}, ignoring all cold and hot nuclear matter effects. The HFE production in \pp\ collisions is evaluated using PYTHIA8 STAR heavy flavor tune~\cite{ref:pythia8, ref:jpsipppau} for $\sqrt{s_{\rm NN}} \leq$ 200 GeV and the upper limits of fixed-order next-to-leading logarithm (FONLL) calculations~\cite{ref:QCDCB} for $\sqrt{s_{\rm NN}} $ = 2.76 and 5.02 TeV as preferred by data~\cite{ALICE:NPE, ALICE:NPE2}.

\section{Results}

\begin{figure}[!htb]
\centering
\includegraphics[width=\linewidth]{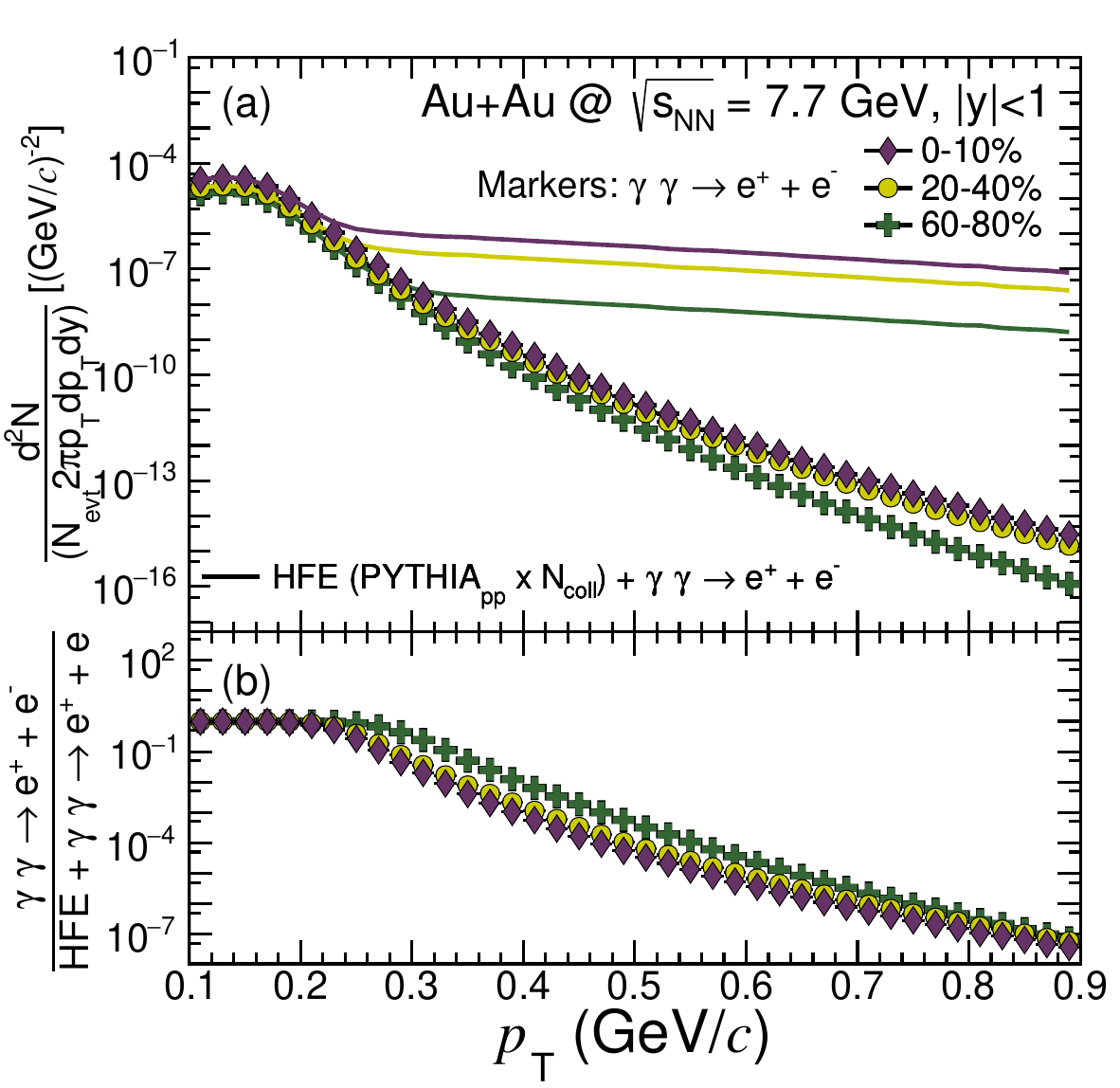}
\includegraphics[width=\linewidth]{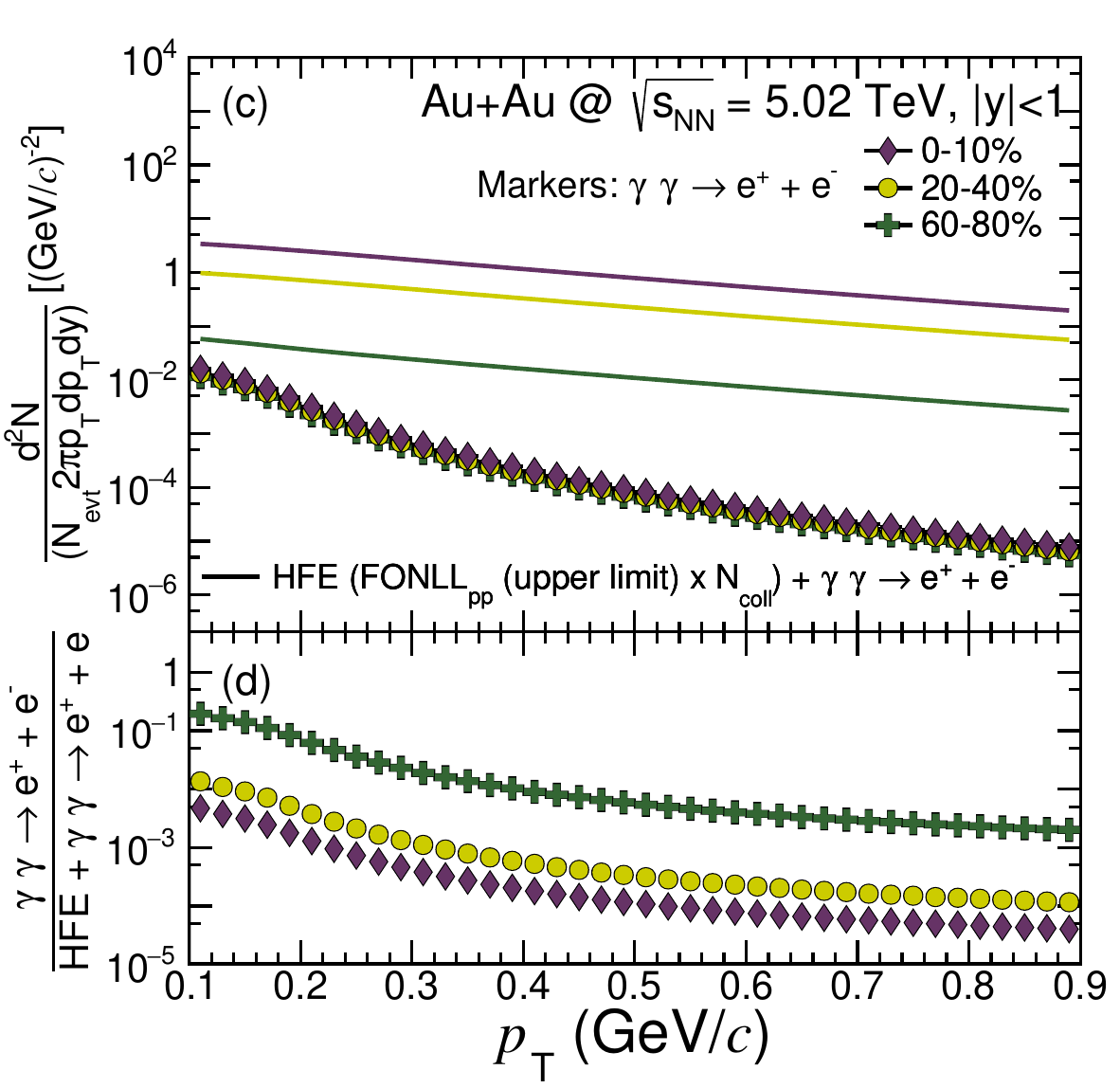}
\vspace{-2ex}
\caption{(a) Invariant yields of coherently produced electrons (markers) for different centrality classes of Au+Au collisions at \sNN{7.7} GeV, along with the sum of the HFE yields and the coherent production (curves). (b) Ratios of coherent production to the sum. (c) (d) Same as (a) and (b) except that it is for 5.02 TeV.}
\label{fig2}
\end{figure}

Invariant yields of electrons from coherent photon-photon interactions are shown in Fig. \ref{fig2} as a function of $\pt$ for different centrality classes (0-10\%, 20-40\%, 60-80\%) of Au+Au collisions at \sNN{7.7} GeV and 5.02 TeV, respectively. The spectra fall steeply with increasing $\pt$. When summed up with the HFE production, the combined yields are shown as solid curves in the figure, which are significantly above the coherent production above 0.3 (0.1) \gevc\ for collisions of 7.7 GeV (5.02 TeV). Fractions of coherently produced electrons in the summed yields as a function of $\pt$ are also shown in Fig. \ref{fig2}. For Au+Au collisions at \sNN{7.7} GeV, the coherent production dominates over the HFE production for $\pt < 0.25$ \gevc, while the fraction quickly drops down to 0.1\% at $\pt = 0.5$ \gevc. On the other hand, for collisions at \sNN{5.02} TeV, the contribution of coherent production is subdominant throughout the examined range of $0.1 < \pt < 0.9$ \gevc, with a strong centrality dependence. The fraction varies between 19.5\%-0.2\% for 60-80\% peripheral collisions, while it is almost always less than 1\% in 0-10\% and 20-40\% central collisions. In Pb+Pb collisions, similar behaviors are observed, as shown in Fig. \ref{fig5}.

\begin{figure}[!htb]
\centering
\includegraphics[width=\linewidth]{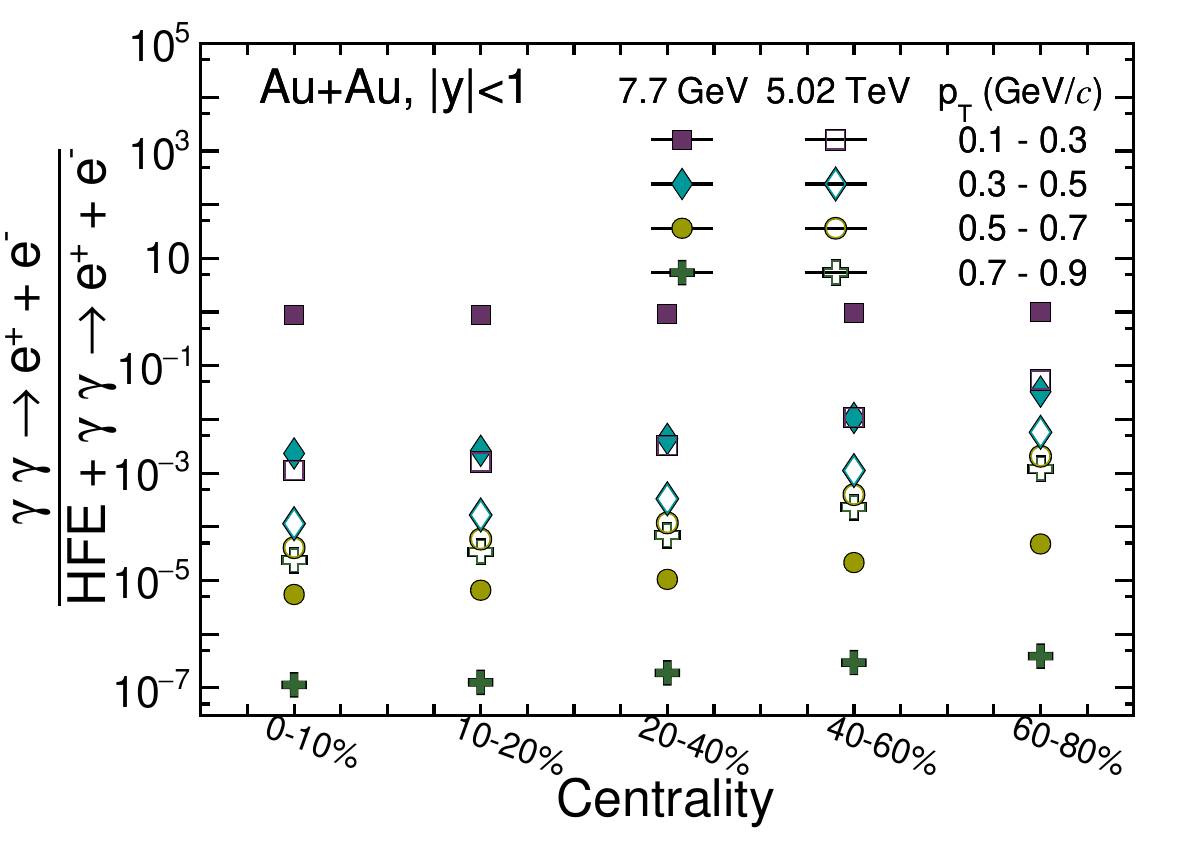}
\vspace{-2ex}
\caption{Ratios of coherently produced electron yields over the sum of the HFE yields and the coherent production as a function of centrality for different $\pt$ ranges in Au+Au collisions at \sNN{7.7} GeV and 5.02 TeV.}
\label{fig3}
\end{figure}

Figure~\ref{fig3} shows the contribution of coherently produced electrons to the sum of HFE yields with the coherent production as a function of centrality in 7.7 GeV and 5.02 TeV Au+Au collisions. The four different $\pt$ ranges from 0.1-0.3 \gevc\ to 0.7-0.9 \gevc\ are compared, and a clear hierarchy is seen. For example, the fraction drops from above 87\% for 0.1-0.3 \gevc\ to about $10^{-7}$ for 0.7-0.9 \gevc\ at 7.7 GeV, which confirms the strong $\pt$ dependence seen in Fig. \ref{fig2}. In all $\pt$ ranges, a clear centrality dependence of the fraction increasing from central to peripheral collisions is observed. For instance, the ratio rises from 87\% (0.2\%) in 0-10\% central collisions to 99\% (3.3\%) in 60-80\% peripheral collisions for 0.1-0.3 \gevc\ (0.3-0.5 \gevc) in 7.7 GeV Au+Au collisions. There is a much stronger centrality dependence, but a much weaker $\pt$ dependence for the ratio at 5.02 TeV compared to that at 7.7 GeV. Similar dependence is seen for Pb+Pb collisions, as shown in Fig. \ref{fig6}.

\begin{figure}
\centering
\includegraphics[width=\linewidth]{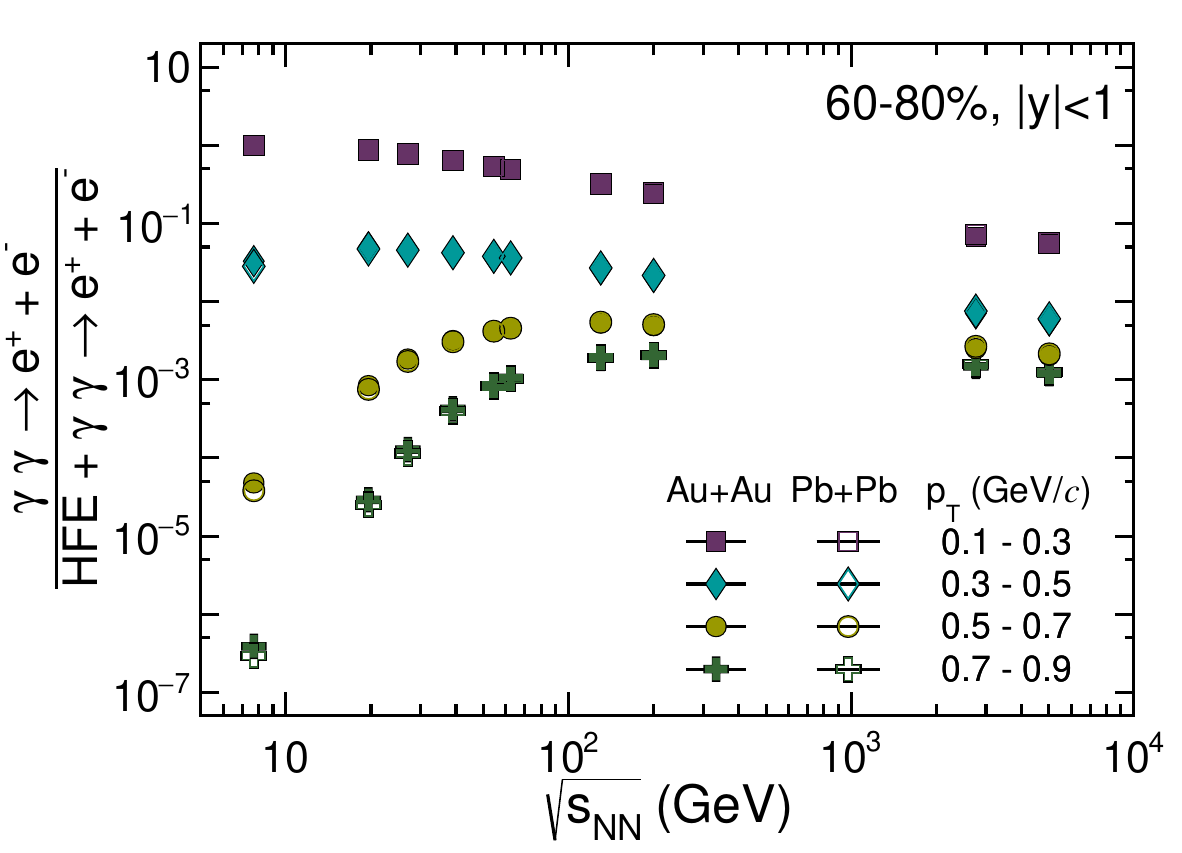}
\vspace{-2ex} 
\caption{Ratios of coherently produced electron yields over the sum of the HFE yields and the coherent production as a function of $\sqrt{s_{\rm NN}}$ for different $\pt$ ranges in 60-80\% Au+Au (solid markers) and Pb+Pb (open markers) collisions.}
\label{fig4}
\end{figure}

In Fig. \ref{fig4}, ratios of coherent production to the sum of coherent production and HFE are shown as a function of collision energy in different $\pt$ ranges for 60-80\% Au+Au and Pb+Pb collisions. Within $0.1 < \pt < 0.3$ \gevc, the fraction decreases monotonically from 7.7 GeV to 5.02 TeV. For the other three $\pt$ ranges, the fraction first increases with energy and then decreases, with the turning point moving to higher collision energy for electrons of higher $\pt$. This is caused by the fact that the different $\pt$ ranges correspond to different \ee\ pair masses. This also leads to a stronger electron $\pt$ dependence of the ratio with decreasing collision energy. For $\pt > 0.3$ \gevc, the ratios are seen to be lower in Pb+Pb collisions than in Au+Au collisions for $\sqrt{s_{\rm NN}} \leq$ 62.4 GeV, while the order reverses at $\sqrt{s_{\rm NN}} >$ 62.4 GeV. This is due to the interplay of the peaked structure of the photon flux (Fig. \ref{fig1new}) and smaller impact parameter values in Au+Au collisions than Pb+Pb collisions (Tab. \ref{tab1}).


\section{Summary}

The recently discovered coherent electron production in hadronic heavy-ion collisions prompted the current quest of its impact as a background source on the measurements of heavy flavor decayed electrons. Thus, we perform calculations of the electron production at mid-rapidity ($|y|<1$) from coherent photon–photon interactions through the Breit-Wheeler process in Au+Au and Pb+Pb collisions. The ratios of these coherently produced electrons relative to the sum of HFEs and the coherent production are reported as a function of electron $\pt$, centrality and collision energy. We found the relative yield of coherent production over HFEs is substantial at low electron $\pt$ and low collision energy. This calls for the need to take it into account for HFE measurements in these kinematic ranges when studying for example cold nuclear matter effects or total charm quark production cross section. The effect drops quickly at higher electron $\pt$ or higher collision energy. In the future, we plan to expand such calculations to investigate the impact of coherently produced electrons on the measurement of collective flow for HFEs. 

\begin{acknowledgements}
This work was supported by National Key R\&D Program of China with Grant No. 2022YFA1604900, National Natural Science Foundation of China with Grant No. 12305145, No. 12175223 and No. 12347101. R. Ma is supported by the U.S. DOE Office of Science under contract Nos. DE-SC0012704, DE-FG02-10ER41666 and DE-AC02-98CH10886. Q. Yang is supported by Natural Science Foundation of Shandong Province of China with Grant No. ZR2020QA086. W. Zha is supported by Anhui Provincial Natural Science Foundation No. 2208085J23 and Youth Innovation Promotion Association of Chinese Academy of Sciences.
\end{acknowledgements}

\bibliographystyle{spphys}
\bibliography{svjourn3-epjc}

\section*{APPENDIX}
\begin{figure}[htb]
\centering
\includegraphics[width=\linewidth]{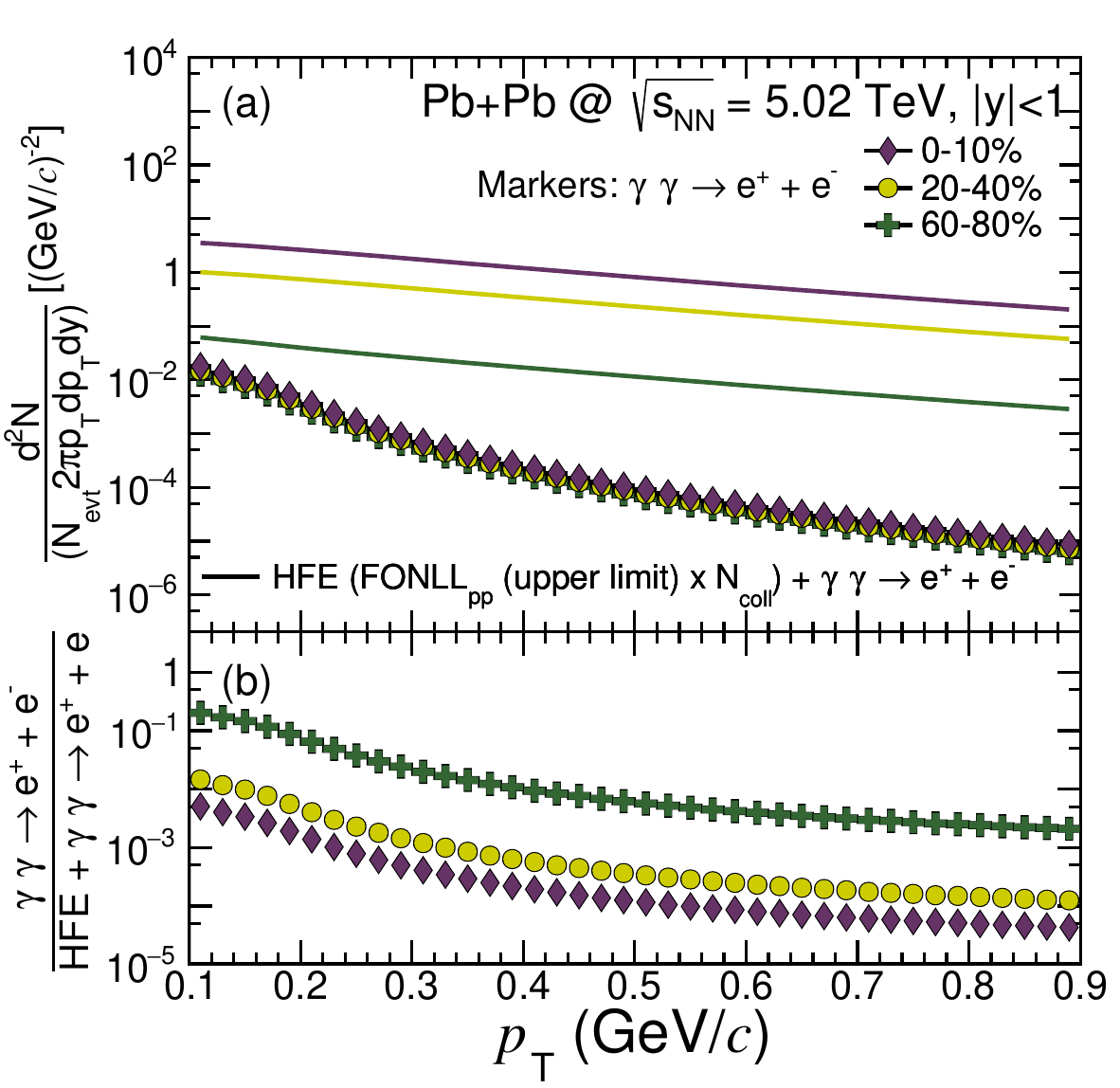}
\vspace{-2ex}
\caption{(a) Invariant yields of coherently produced electrons (markers) for different centrality classes of Pb+Pb collisions at \sNN{5.02} TeV, along with the sum of the HFE yields and the coherent production (curves). (b) Ratios of coherent production to the sum.}
\label{fig5}
\end{figure}

Figures \ref{fig5} (a) shows the invariant yields of electrons from coherent photon-photon interactions as a function of $\pt$, along with the sum of HFE yields with the coherent production, for different centrality classes (0-10\%, 20-40\%, 60-80\%) of Pb+Pb collisions at \sNN{5.02} TeV. The panel (b) shows the fractions of coherently produced electrons in the summed yields as a function of $\pt$. 

\begin{figure}[!htb]
\centering
\includegraphics[width=\linewidth]{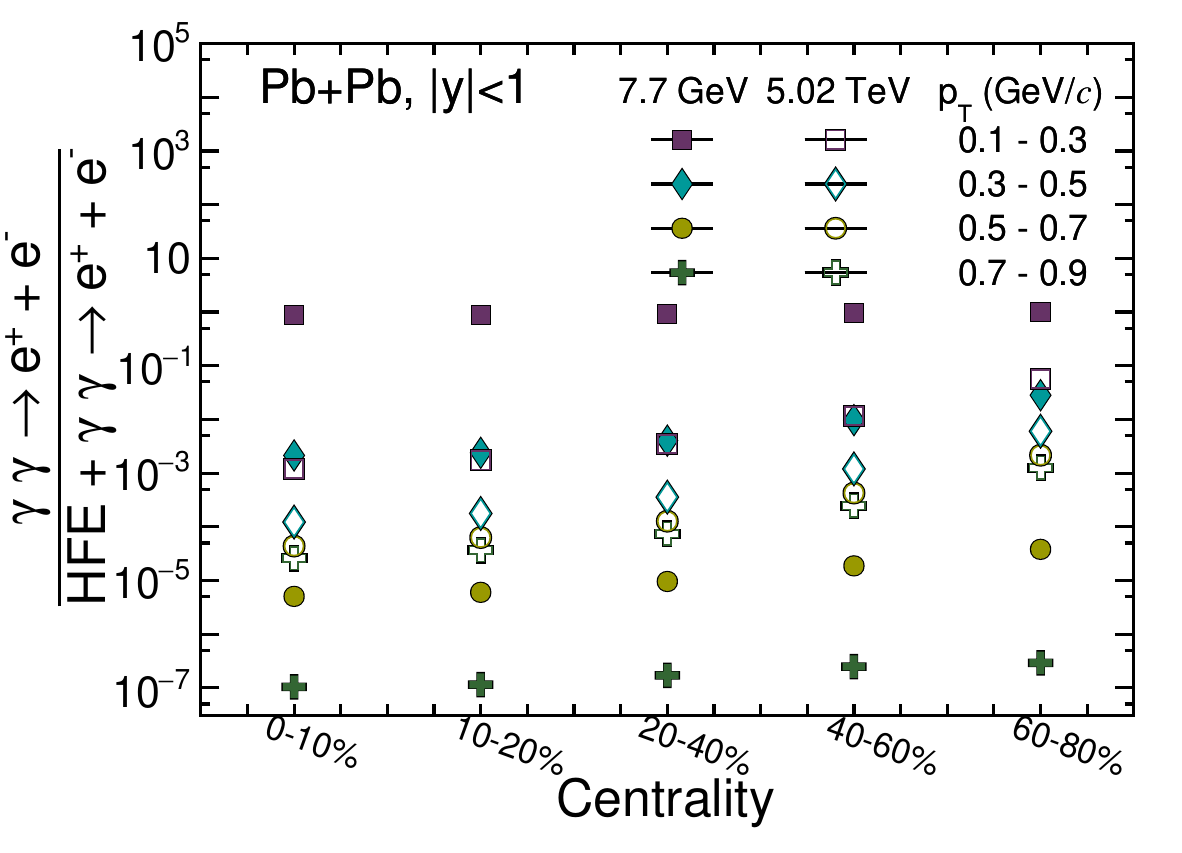}
\vspace{-2ex}
\caption{Ratios of coherently produced electron yields over the sum of the HFE yields and the coherent production as a function of centrality for different $\pt$ ranges in Pb+Pb collisions at \sNN{7.7} GeV and 5.02 TeV.}
\label{fig6}
\end{figure}

Figure ~\ref{fig6} shows the contribution of coherently produced electrons to the sum of HFE yields with the coherent production as a function of centrality in 7.7 GeV and 5.02 TeV Pb+Pb collisions. 

\end{document}